# Phase transformations in metastable β Zr15Nb alloy revealed by in-situ methods

Anna Veverková[a,*], Kristína Bartha[a], Jozef Veselý[a], Pere Barriobero-Vila[b,c], Jiří Kozlík[a], Petr Doležal[d], Jiří Pospíšil[d], Jana Šmilauerová[a], Josef Stráský[a]

[a] Department of Physics of Materials, Charles University, Ke Karlovu 5, 121 16 Prague, Czech Republic
[b] Department of Materials Science and Engineering, Technical University of Catalonia-BarcelonaTech (UPC), Eduard Maristany 16, 08019 Barcelona, Spain
[c] CIM UPC, Llorens i Artigas 12, 08028 Barcelona, Spain
[d] Department of Physics of Condensed Matter, Charles University, Ke Karlovu 5, 121 16 Prague

* Corresponding author. E-mail address: veverkova@karlov.mff.cuni.cz

**Abstract**

This study examines the phase transitions occurring during linear heating of the Zr15Nb alloy through a comprehensive, multi-technique methodology comprising *in-situ* high-energy synchrotron X-ray diffraction (HEXRD), electrical resistance measurements, differential scanning calorimetry (DSC), and thermal expansion analysis, supplemented by *ex-situ* transmission electron microscopy (TEM). The findings reveal a complex sequence of phase transformations and corresponding structural changes over a broad temperature range (from room temperature up to 800 °C). Two distinct body-centered cubic (bcc) β phases—$β_{Zr}$ and $β_{Nb}$—with closely related lattice parameters are identified. At room temperature, the microstructure is characterized by a mixture of the metastable $β_{Zr}$ + $ω_{ath}$ phase. Upon heating, $β_{Zr}$ progressively decomposes, giving rise to the formation of $β_{Nb}$. TEM observation revealed the cuboidal shape of the $ω_{iso}$ particles resulting from the high lattice misfit between β and ω phase. The ω solvus temperature is determined to be approximately 555 °C, as evidenced by in situ HEXRD and abrupt changes in the alloy's thermal and electrical properties. The growth of the α phase occurs after the dissolution of the ω phase, resulting in a pronounced increase in thermal expansion.

1. Introduction

Zirconium possesses a range of properties that render it suitable for various industrial applications. It is utilized in the chemical industry [1] and considered for utilization in biomedicine due to its biocompatibility and low elastic modulus [2]. However, zirconium is most extensively employed in the nuclear industry, particularly for nuclear fuel cladding, owing to its outstandingly low thermal neutron capture cross section [3, 4]. Utilization of Zr alloys at elevated temperatures above 500 °C is limited due to deterioration of mechanical properties and accelerated oxidation.

Mechanical properties of Zr, like in any other metal, can be significantly improved by alloying [5]. However, a notable disadvantage of alloying is the increase in the overall thermal neutron capture cross section due to presence of alloying elements [6]. Therefore, in most cases, zirconium alloys with a very low content of alloying elements have been designed, studied, and used [6–12]. Main alloying elements in Zr alloys include Sn, Fe, Nb, or Mo.

Less frequent studies, however, show that zirconium alloys with higher alloying can exhibit excellent strength and mechanical properties at room temperature and, in particular, at elevated temperatures [13–17]. The development of such stronger alloys would enable a subtler design of the nuclear fuel cladding which could compensate for the adverse effect of the increased cross section for neutron capture. Additionally, this subtler design would reduce component weight and production costs [4, 6, 18].

Zirconium alloys with a higher content of alloying elements were studied sporadically. Near β-type alloys were studied for biomedical use [15, 17], the Zr-Nb system was considered for use in magnetic resonance imaging [19, 20]. The lack of reports on high-alloyed Zr alloys is particularly striking when compared with the decades of development devoted to Ti alloys, which share many similarities as members of the same group in the periodic table. Specifically, Ti–Nb alloys and their phase transformations have been extensively studied [21–23].

To our knowledge, the sequence of phase transformations in metastable β-Zr alloys was investigated only in our previous study for Zr12Nb alloy [24].

Quenching of a Zr12Nb metastable β alloy from a temperature above the β transus temperature results in material containing the high-temperature β phase (body-centered cubic) "frozen" in a metastable condition. In the metastable β matrix, particles of the metastable $\omega_{ath}$ (athermal) phase are formed during quenching. The mechanism of ω phase formation has been described in plentiful studies [25–27]. As the temperature increases, diffusion starts to play a significant role and the ω phase rejects β-stabilizing elements (i.e. Nb in this case) into the surrounding β matrix; this chemically stabilized ω is denoted as $\omega_{iso}$ (isothermal). At the same time, as the niobium content in the β matrix is enhanced and it decomposes to two bcc structures with different lattice parameter $\beta_{Zr}$ phase and $\beta_{Nb}$ phase form. Such effect is not observed in most Ti alloys. At around 560 °C, the ω phase dissolves and the α phase starts to precipitate. Simultaneously, the zirconium rich $\beta_{Zr}$ phase disappears. As the alloy approaches the temperature of the β transus, α phase gradually dissolves. Above the β transus, the material consists of pure bcc matrix, referred again to as $\beta_{Zr}$ phase [24].

As follows from the binary phase diagram calculated in Thermo-Calc software (TTZR1: Zr-based alloys database v1.1) shown in Figure 1, the phase stability strongly depends on the Nb content. In this study, Zr15wt%Nb alloy was studied, i.e. with 3 wt% Nb more than in our previous study. The degree of β stabilization is similar to the thoroughly studied Ti15wt%Mo alloy, and allows for direct comparison [28].

This experimental study aims on determining the effect of increased Nb on undergoing phase transformations, namely on suppression of ω phase formation, its solvus temperature and also on $\beta_{Zr}$ and $\beta_{Nb}$ separation [24].

A suitable method for investigating phase transformations *ex-situ* and *in-situ* is X-ray diffraction (XRD). However, laboratory X-ray diffraction provides only limited penetration depth and, most importantly, long acquisition times that are not appropriate to determine the ongoing phase transformations *in-situ* during linear heating. These disadvantages can be overcome by employing high-energy X-ray diffraction (HEXRD) at a synchrotron facility. The method has already been extensively used for detailed studies of phase transformations in titanium alloys [28–31] and, as mentioned in detail above, for the Zr12Nb alloy [24]. In this study, HEXRD data are carefully investigated to reveal in detail two distinct features of phase transitions sequence: coexistence of β + ω + α phase and diminishing of $\beta_{Zr}/\beta_{Nb}$ separation along with $\omega_{iso}$ phase dissolution, which seems to be a unique process observed only in Zr alloys.

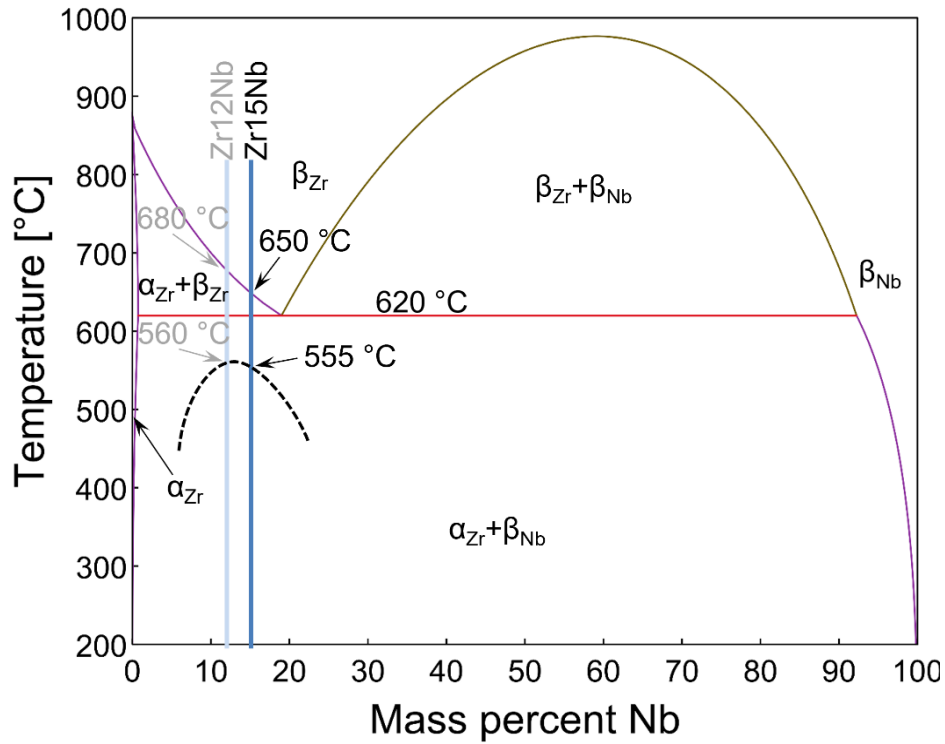

**Fig. 1:** Binary phase diagram of the Zr-Nb system calculated in Thermo-Calc software (TTZR1: Zr-based alloys database v1.1). The light blue and dark blue vertical lines denote the compositions of the Zr12Nb alloy (studied previously in [24]) and the Zr15Nb alloy (present study), respectively. The dashed line schematically shows the variation of ω solvus temperature with Nb content. The ω solvus temperature of the Zr15Nb alloy (555 °C) was determined in this study.

## 2. Experimental methods

The studied Zr15Nb alloy was arc melted in He atmosphere at UJP Praha a.s., homogenized at 1400 °C for 2 hours in vacuum and furnace cooled. Subsequently, the solution treatment (ST) was performed at 1000 °C for 2 hours in vacuum and immediately water quenched.

The concentrations of oxygen, nitrogen and hydrogen in the alloy were determined using carrier gas hot extraction (CGHE). Triplicate measurements were performed, yielding an average oxygen, nitrogen and hydrogen content of *(991 ± 55) wppm*, *(815 ± 22) wppm* and *(75 ± 2) wppm*, respectively.

To study phase transformations, high-energy synchrotron X-ray diffraction (HEXRD) was performed during linear heating from room temperature (RT) up to 800 °C with 5 °C/min heating rate at the P07-HEMS beamline of PETRA III (Deutsches Elektronen-Synchrotron) [32]. A modified Bähr 805A/D dilatometer was used to heat the sample in He atmosphere and measure the thermal expansion simultaneously [33]. The measurement was carried out in transmission mode on a cylindrical sample with a diameter of 4.5 mm and a length of 10 mm. The primary beam with a wavelength of 0.14235 Å was perpendicular to the length of the specimen. The patterns of entire Debye-Scherrer rings were recorded by a PerkinElmer XRD 1621 image plate detector. One pattern consisted of the sum of 15 frames, each with an acquisition time of 1 s. The distance between the detector and the sample was about 1580 mm and the beam size was 0.7 mm x 0.7 mm. The obtained Debye-Scherrer rings were processed by the software fit2D to get integrated 2θ profiles [34]. Subsequently, the profiles were refined using the Rietveld method in the FullProf program [35]. The detection limit of the HEXRD method depends on the volume fraction, the size of particles associated with a given phase, separation of peaks from other phases, in the particular of the ω phase, and also on the degree of plane collapse [36].

Complementary techniques, namely electrical resistance measurement and differential scanning calorimetry (DSC), were employed in a manner analogous to HEXRD, i.e. *in-situ* during linear heating from RT to 800 °C at a constant rate of 5 °C/min. All measurements were conducted under an inert argon atmosphere. To ensure the repeatability, two specimens were measured for each condition. Electrical resistance was measured using the four-point probe method with a custom-built apparatus, as described in [37], and the resulting values were normalized to the resistance at RT. DSC data were obtained using a Netzsch DSC 404C Pegasus calorimeter.

TEM microstructure observations were performed *ex-situ* on pre-annealed samples. For this purpose, ST samples were encapsulated in a quartz tube filled with Ar, linearly heated at a heating rate of 5 °C/min, and quenched into water with immediate breaking of the capsule. Two-step polishing was applied after mechanical grinding: electropolishing in Tenupol-5 (Struers) in a solution of $HClO_4$, butanol and methanol at -20 °C using 20 V, and final ion milling in PIPS (Gatan) at 5 kV. In the last 30 min of ion milling, the voltage was reduced to 3 kV to ensure a smooth surface of the TEM observed area. For TEM observations, a JEOL 2200FS transmission electron microscope operated at 200 keV was used.

## 3. Results

### 3.1. In-situ high energy X-ray diffraction (HEXRD)

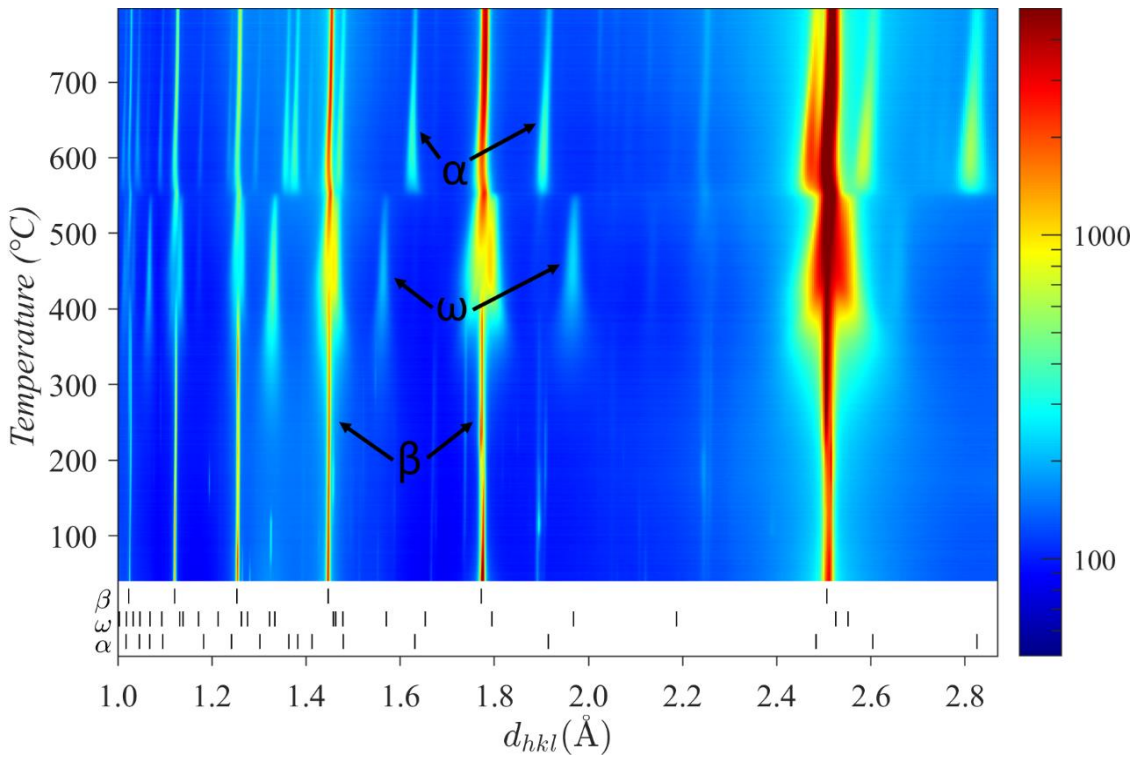

**Fig. 2:** Evolution of the X-ray diffraction patterns of the Zr15Nb alloy during linear heating from RT to 800 °C. The interplanar distance is shown on the horizontal axis, while the temperature is plotted on the vertical axis. The color scale represents the diffracted intensity on a logarithmic scale. The lower part shows the positions of the peaks of β, ω and α phases

Figure 2 shows the temperature evolution of the diffracted intensity of the Zr15Nb alloy during linear heating from RT to 800 °C at a heating rate of 5 °C/min. The positions of the peaks of β, ω and α phases are indicated in the lower part of the figure. Note that although two types of the β phase ($β_{Zr}$ and $β_{Nb}$) are present in the material at certain temperatures, only one β phase is labeled in Figure 2 due to adjacent peaks. The difference between $β_{Zr}$ and $β_{Nb}$ will be discussed below.

At RT, only β peaks are visible in the diffraction patterns (e.g. 1.45 Å, 1.77 Å, 2.51 Å). Nevertheless, it was shown by TEM (see below, Figure 6) that the material consists of a mixture of β + ω phase at RT. However, up to approximately 250 °C, both the amount of the ω phase and the size of ω phase particles are below the detection limit of the HEXRD. Above 250 °C, the peaks of the ω phase appear (e.g. at 1.33 Å (double peak), 1.57 Å and 1.97 Å) and sharpen with increasing temperature. Around 400 °C an apparent broadening of the β peaks is visible – the reasons of the broadening are clarified in the Discussion section.

At 555 °C, an abrupt change corresponding to the dissolution of the ω phase at the ω solvus temperature is observed [24]. Concurrently, the β peaks sharpen and slightly shift to lower values of the interplanar distance (this change in the lattice parameter is caused by diffusion of Nb from evolving α phase). Simultaneously, around 555 °C, the α phase starts to precipitate. Significant peaks of the α phase are well visible e.g. at 1.63 Å or 1.91 Å. Finally, as the temperature approaches 800 °C (upper limit of the HEXRD experiment), the intensity of the α peaks weakens as it dissolves in the vicinity of the β transus temperature.

To analyze properly the phase evolution taking place during linear heating of the material, the HEXRD patterns were fitted using the Le Bail method implemented in the FullProf program [35]. The evolutions of the lattice parameters and hence molar volumes of individual phases were obtained from the fit. Unfortunately, due to large grain size and consequently poor statistics of the diffraction data, it was not possible to evaluate unambiguously the volume fractions of the phases. Besides the presence of α and ω phases, two types of bcc phases ($β_{Zr}$ and $β_{Nb}$) with different lattice parameters were present in the material during linear heating. The temperature evolution of the lattice parameters of these four phases ($β_{Zr}$, $β_{Nb}$, ω and α) during linear heating is shown in Figure 3. The lattice parameter of both β phases $a_β$ is plotted to show the lattice misfit between ω and β phases i.e. multiplied by √2 and √3/2, respectively [27].

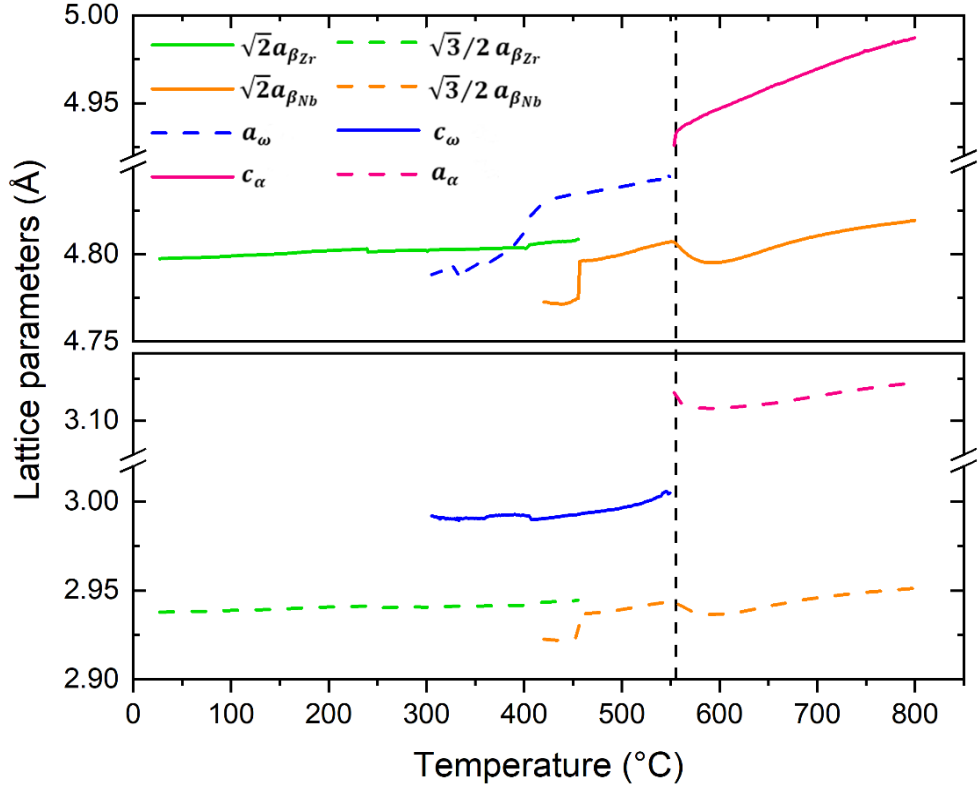

**Fig. 3:** Evolution of the lattice parameters of individual phases during linear heating. The lattice parameter $a_\beta$ is plotted to show the lattice misfit between ω and β phases as defined by Eq. (1), i.e. multiplied by √2 and √3/2. The vertical dotted line represents the temperature of ω solvus at 555 °C

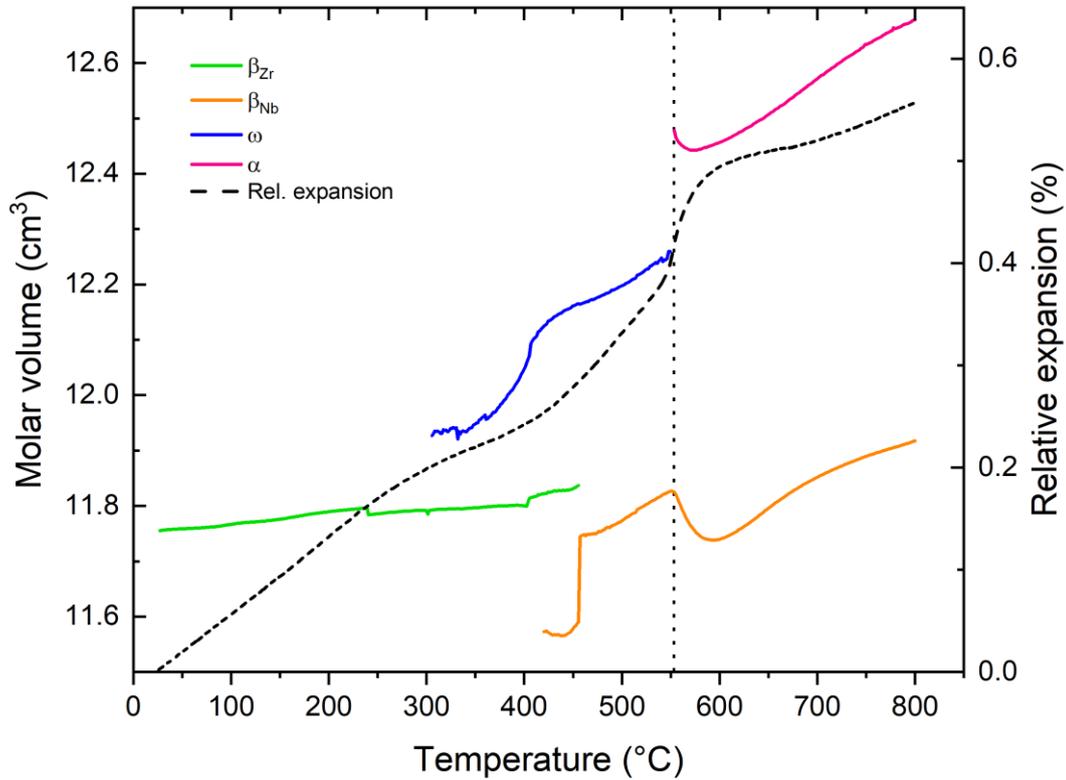

**Fig. 4:** Evolution of the molar volume of individual phases during linear heating. The relative change of the sample length (black dashed line) is plotted in the same graph. The vertical dotted line represents the temperature of 555 °C.

Figure 4 shows the evolution of the molar volume of all the detected phases (colored lines) during heating and of the relative change in the sample length measured simultaneously with HEXRD (black dashed line). In an ideal crystal, the crystal lattice expands linearly with increasing temperature, and therefore, both the molar volume and the sample length grow linearly. Any deviation from linearity corresponds to ongoing processes in the material. In our case (Figure 4), we observe several deviations from the linear trends, which will be explained in the Discussion.

Note that the jump in both the lattice parameter and the molar volume of $β_{Zr}$ (Figures 3 and 4, respectively) is a fitting artifact resulting from the change of the number of phases in the refinement.

3.2. Differential scanning calorimetry and electrical resistance measurement

To obtain more information on the ongoing phase transformations, the evolution of the heat flow and electrical resistance upon heating (the same heating regime as for HEXRD) was measured. Figure 5 shows the electrical resistance (black line), derivative of the electrical resistance (gray line), and the heat flow (red line) measured during linear heating of the Zr15Nb alloy. The derivative of the relative change of the sample length is also plotted (green line) as it can be directly correlated to the DSC curve.

In an ideal case, a DSC (heat flow) curve of a material which does not undergo any transformation is a constant line. Any deviation from the constant baseline reflects the occurrence of some process in the material. In general, the first-order transformations manifest as peaks (exothermic or endothermic) due to latent heat effects, while the second-order transformations appear as steps in the baseline due to heat capacity changes.

In the following, the heat flow curve is presented with exothermic events plotted in the positive direction and endothermic events in the negative one. The present Zr15Nb measurement shows a broad and relatively shallow exothermic event beginning at approximately 200 °C. Superimposed on this effect is a more distinct exothermic peak occurring between 350 °C and 400 °C. This is immediately followed by an endothermic process between approximately 420 °C and 550 °C. Finally, a sharp exothermic peak is observed between 550 °C and 650 °C. These thermal events are analyzed in detail in the Discussion section.

According to Matthiessen's rule, the electrical resistivity of a metal at a given temperature $T$ can be determined as follows:

$$\rho_{TOT}(T) = \rho_h(T) + \rho_0, \tag{3}$$

where the temperature-dependent part $\rho_h(T)$ is the resistivity of a metal with an ideal lattice structure. $\rho_h(T)$ is given mainly by phonon scattering and increases linearly with temperature. This part is responsible for an increase of the resistivity by approximately 20 % [38]. The temperature-independent $\rho_0$ part is the contribution of lattice defects such as foreign atoms, phases, phase interfaces or strains [39]. Therefore, deviations from the linear trend in a resistivity measurement correspond to some ongoing processes in the material. As the specific value of electrical resistivity is not relevant (it depends on sample geometry), we plot only the relative change of the electrical resistance, i.e. the electrical resistance normed to the value at room temperature.

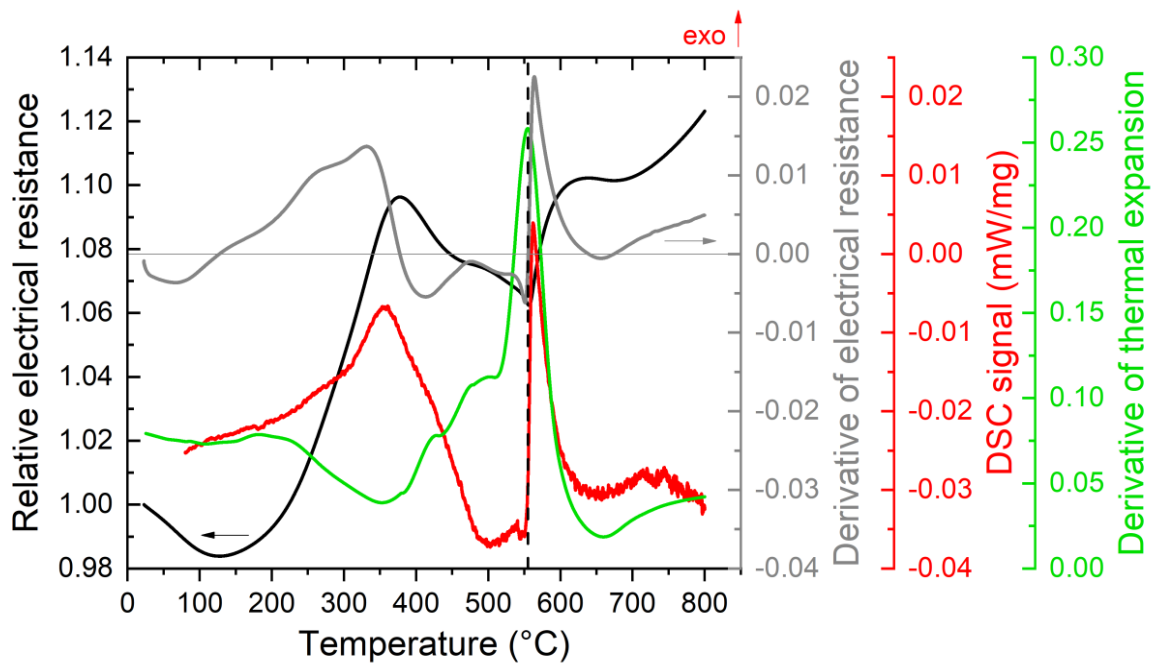

**Fig. 5:** The temperature evolution of the relative electrical resistance (black line), its derivative (gray line), DSC (red) and the derivative of thermal expansion (green) of the Zr15Nb alloy during linear heating. The grey horizontal line indicates zero value of the derivative of electrical resistance. The direction of the exothermic processes is indicated

3.3. Ex-situ observation of the microstructure

The microstructure was observed *ex-situ* using TEM. Four temperatures were chosen according to the HEXRD results (400 °C, 500 °C, 540 °C, and 600 °C) and the samples were prepared by linear heating up to these temperatures and subsequent water quenching. TEM was performed on these *ex-situ* samples as well as on the ST sample. Unfortunately, *in-situ* studies of phase transformations by TEM turned out to be unfeasible, as the driving forces of phase transformations in a 2D material (i.e. thin TEM foil) are different than in bulk and some phase transformations may occur at a different temperature or may not occur at all in a thin foil [28].

Figure 6 shows the phase compositions of the studied *ex-situ* samples (i.e. back at room temperature) which may differ from those present at high temperatures, as some phase transformations might have occurred during quenching. For instance, $\omega_{ath}$ may form during quenching as explained in the next lines for Figure 6f.

In Figure 6a), a dark-field (DF) image of the ST sample is shown. The tiny bright spots correspond to ω-phase particles which satisfy the diffraction condition. As the specimen is slightly bent, the ω-phase particles are visible only in regions of the sample in which the diffraction condition is fulfilled. Note that the presence of tiny ω phase particles (size of several nanometers) in the ST sample was proven by TEM, although the ω phase was not detected by HEXRD at RT (see Figure 2).

DF micrographs of samples pre-heated to 400 °C, 500 °C and 540 °C are shown in Figures 6b), c) and d), respectively. During heating to 400 °C, the size of the ω particles slightly increases (Figure 6b)). With further heating, ω particles grow and they are significantly larger (tens of nanometers) at 500 °C (Figure 6c)) and even larger at 540 °C (Figure 6d)) and their shape is cuboidal. Note that the number of particles cannot be determined from TEM observation; even though the exact thickness of the observed region could be determined, only ω-phase particles fulfilling the diffraction condition are observed and the quantitative determination would be misleading.

Figures 6 e) and f) show the microstructure of the sample heated to 600 °C. As the material consists of a mixture of α+β phase, both bright-field (BF) and dark-field (DF) images from the same sample area are displayed. In the BF image, Fig 6e), the presence of α particles is shown, while the DF micrograph (Figure 6f) shows ω phase particles in the remaining β matrix (between α particles). These ω phase particles have significantly smaller size compared to stabilized $\omega_{iso}$ particles observed at 540°C in Figure 6d). This is caused by the fact that the temperature of 600 °C lies above the ω solvus, at which the stabilized $\omega_{iso}$ particles disappeared during linear heating. Small ω particles can be therefore identified as $\omega_{ath}$ formed during quenching from 600 °C.

Figures 6g) and h) show two types of selected area electron diffraction (SAED) patterns corresponding to the DF micrographs presented above. Figure 6g) displays a SAED pattern of the ST sample in orientation $(110)_\beta$. The sample preheated to 540 °C was observed in the same orientation (Figure 6d). The SAED pattern in Figure 6h) belongs to the sample pre-heated to 400 °C in orientation $(113)_\beta$. The specimens pre-heated to 500 °C and 600 °C (Figures 6c), e), and f)) were also observed in the $(113)_\beta$ orientation. Note that in both orientations, only one family of the ω phase particles is visible. The SAED pattern corresponding to the sample pre-heated to 600 °C contains diffraction spots belonging to all three phases: α, β and ω (the SAED pattern is not presented here).

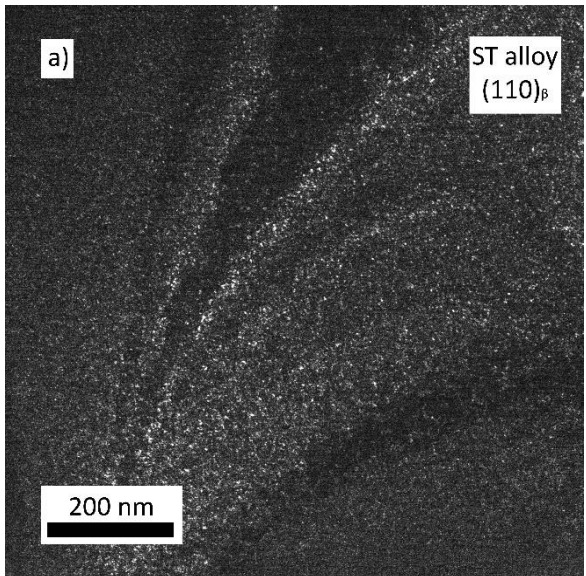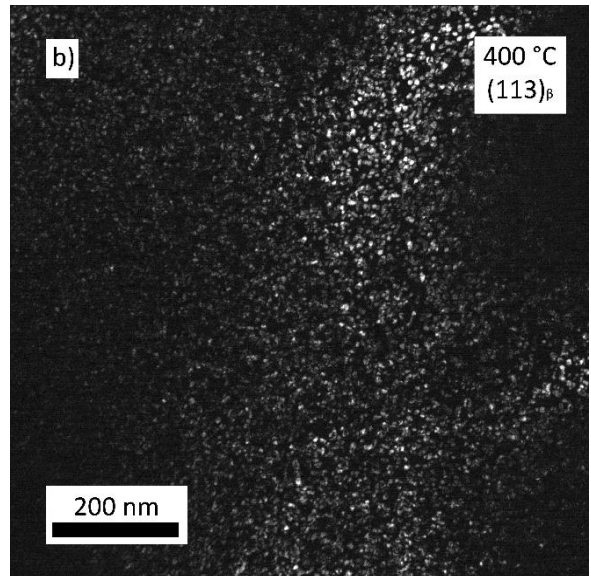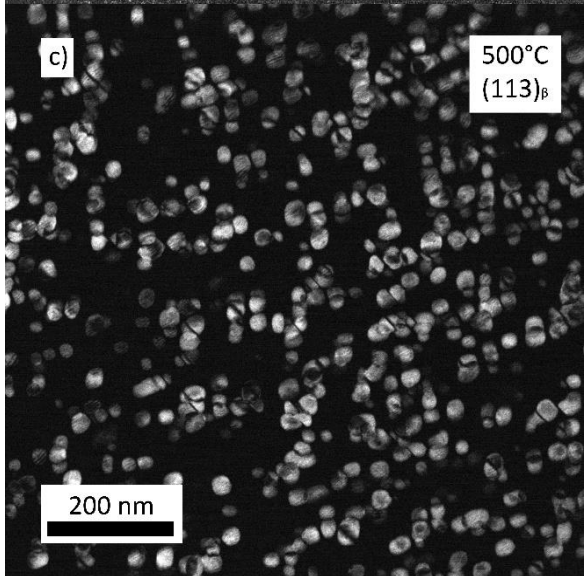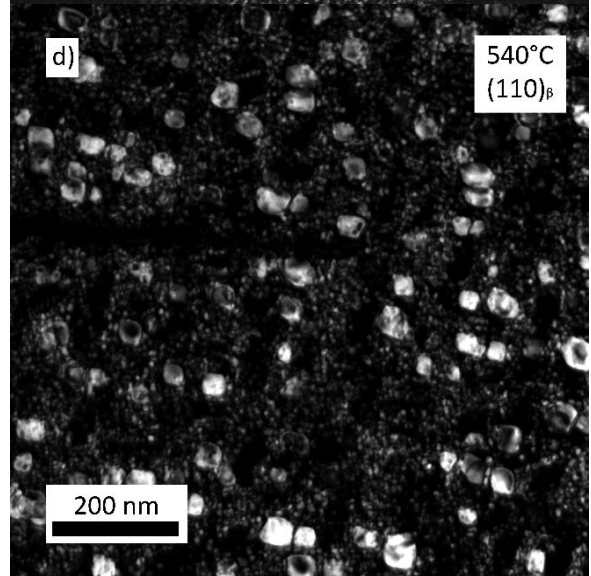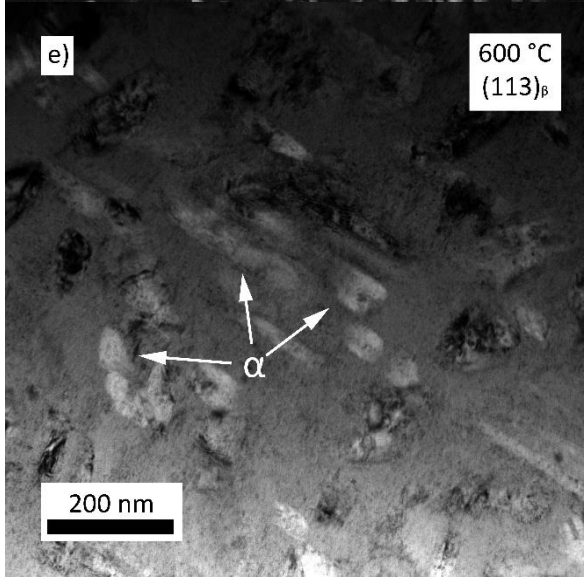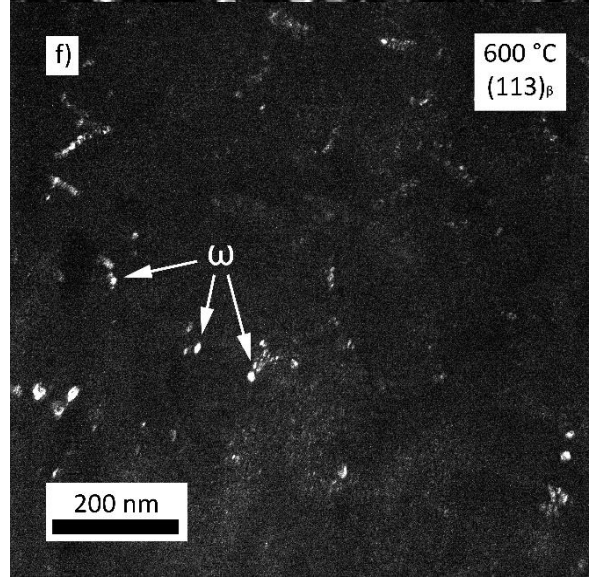

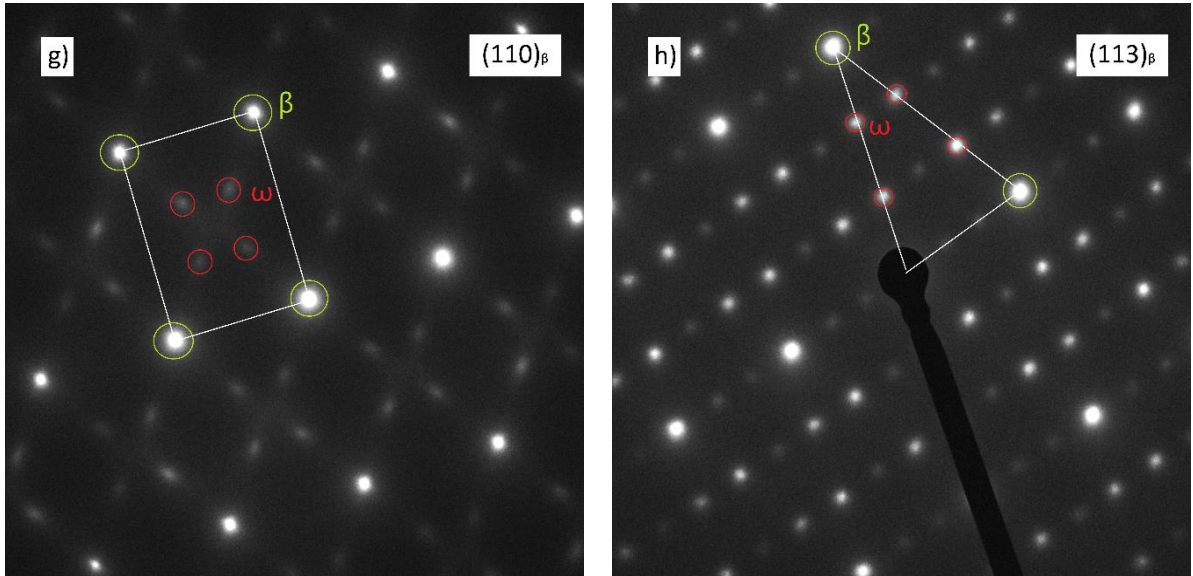

**Fig. 6:** TEM micrographs of the Zr15Nb alloy: a) DF image of the ST sample in $(110)_\beta$ direction; DF images of samples preheated to: b) 400 °C ($(113)_\beta$ direction) c) 500 °C ($(113)_\beta$ direction), d) 540 °C ($(110)_\beta$ direction); e) BF image and f) DF image of the Zr15Nb alloy pre-heated to 600 °C in $(113)_\beta$ direction, both images show the same region of the sample; g) SAED pattern of the ST sample, h) SAED pattern of the sample pre-heated to 400 °C

### 4. Discussion

In the following section, the sequence of phase transformations in the Zr15Nb alloy is described and discussed. Processes occurring during heating in the Zr12Nb and Zr15Nb alloys are compared and discussed separately.

#### 4.1. Room temperature

At RT, the Zr15Nb alloy consists of the β matrix with tiny $\omega_{ath}$ phase particles. The ω phase at RT in Zr15Nb cannot be detected by HEXRD (see Figure 2) due to low volume fraction, small size of the particles and their possibly incomplete collapse [36]. However, TEM observations confirmed the presence of ω at RT also in Zr15Nb. The $\omega_{ath}$ particle size is in the order of few nanometers (Figure 6a)). The size of the $\omega_{ath}$ phase particles is significantly smaller in the Zr15Nb alloy in comparison to the Zr12Nb alloy, cf. Figure 6a) in this study and Figure 8a) in [24]. It can be therefore concluded that higher stabilization of Zr alloy by β-stabilizing Nb results in lower fraction and smaller size of $\omega_{ath}$ phase particles.

#### 4.2. Temperature range from RT to 220 °C

From RT to approximately 130 °C, electrical resistance decreases (see the black curve in Figure 5). In metastable β titanium alloys, the initial decrease in electrical resistance was associated with a partial vanish of the $\omega_{ath}$ phase [40]. For several metastable β titanium alloys, it was shown that all processes below approximately 80 °C connected with $\omega_{ath}$ evolution are reversible with no hysteresis [37, 41].

In the temperature range of approximately 130 °C - 220 °C, the electrical resistance gradually increases, which means that the phonon scattering (which increases with increasing temperature) becomes dominant over the process decreasing the electrical resistance (partial vanishing of $\omega_{ath}$ phase). Nevertheless, processes occurring up to approximately 220 °C have no impact on either DSC signal or thermal expansion (see Figure 5).

#### 4.3. Temperature range from 220 °C to 450 °C

A broad exothermic peak appears between 300 °C and 450 °C with a small bump on its shoulder (220 °C - 300 °C), see the red curve in Figure 5. Very similar characteristics can be observed on the derivative of electrical resistance (gray curve in Figure 5).

The first exothermic process (220 °C - 300 °C) slows down the thermal expansion of the sample. This process most likely involves the start of chemical stabilization of ω particles, i.e. the diffusional $\omega_{ath} \rightarrow \omega_{iso}$ transition during which the alloying element Nb is rejected from ω particles into the β matrix.

As reported in [19, 42], the lattice parameter of the $\beta_{Zr}$ phase decreases with increasing Nb content (for instance during Nb depletion from $\omega_{iso}$ to β matrix), which manifests itself macroscopically in the slower increase of thermal expansion upon heating (its derivative decreases, see the green line in Figure 5). A similar phenomenon was also observed for Ti-Mo alloys, in which increasing Mo content in the matrix lowers the lattice parameter of the β phase and decreases the lattice parameters of the ω phase [43, 44]. In other words, the rejection of Mo from the ω particles into the β matrix reduces the lattice constants of both phases. This can be observed macroscopically in the evolution of thermal expansion during linear heating; sample length even decreases over certain temperature range [45]. The particles of the ω phase are still small and coherent in the range of 220 °C - 300 °C but the chemical inhomogeneities resulting from ω stabilization and slight growth may lead to higher conduction electron scattering and therefore to a faster increase in electrical resistance, see the derivative of electrical resistance in Figure 5 (gray line). The faster increase means that the chemical stabilization of ω particles adds to the contribution of phonon scattering. These indirect observations also support the hypothesis of activation of diffusional $\omega_{ath} \rightarrow \omega_{iso}$ transition in 220 °C - 300 °C temperature range. At approximately 250 °C, ω peaks become finally visible in the HEXRD patterns (see Figure 2), however, the intensity of the ω peaks is insufficient for the Le Bail refinement until approximately 300 °C (see Figures 3 and 4).

The second exothermic process (300 °C - 450 °C) observed in DSC is accompanied by a decrease in electrical resistance. The contribution to the decrease of electrical resistance can be deducted as follows: the resistance curve has an inflection point around 330 °C and its increasing trend slows down until the electrical resistance starts to decrease at around 380 °C. This means that a phase transformation process occurring in the material gradually becomes stronger until it prevails over the contribution of gradually increasing phonon scattering in terms of overall electrical resistance evolution. Concurrently, the increase in sample length with temperature is the lowest at approximately 350 °C, which is the consequence of Nb enrichment of the β matrix and the resulting β lattice parameter decrease as explained above.

Between 350 °C and 420 °C, the rate of thermal expansion gradually increases (see the green line in Figure 5). This increase relates to a steep rise of the ω molar volume (see Figure 4). The TEM image of a sample heated to 400 °C shows that ω particles are visibly larger than in the ST sample (see Figure 6b). Concurrently to the chemical stabilization and growth of the ω phase, β phase decomposition and formation of $\beta_{Nb}$ takes place in the material due to significant Nb enrichment of the β matrix (cf. phase diagram in Figure 1). New $\beta_{Nb}$ peaks start to be discernible from $\beta_{Zr}$ peaks in the fit of the HEXRD data at around 420 °C; however, the real onset of $\beta_{Nb}$ formation occurs most probably at lower temperatures. The coexistence of these two bcc phases ($\beta_{Zr}$ and $\beta_{Nb}$ with a higher and lower lattice parameter, respectively) significantly broadens the bcc (β) peaks shown in Figure 2.

Separation to $\beta_{Zr}$ and $\beta_{Nb}$ phases is of particular interest, and therefore, XRD patterns showing the temperature dependence of the 2 0 0 diffraction peak of the β phase are depicted in Figure 7a. The XRD pattern at the lowest depicted temperature around 300 °C consists of only the 2 0 0 $\beta_{Zr}$ peak. Increasing temperature results in a continuous increase of the 2 0 1 ω peak (1.79 Å) and 2 0 0 $\beta_{Nb}$ peak (1.76 Å). Consequently, three phases $\beta_{Zr}$, $\beta_{Nb}$ and $\omega_{iso}$ coexist at temperatures 390 – 450 °C. The peak of the $\beta_{Zr}$ phase decreases with increasing temperature until it seems to disappear completely at around 450 °C. For the sake of clarity, we note here that upon further temperature increase from 450 to 550 °C, the $\beta_{Nb}$ shifts continuously back to the position (lattice parameter) of the original $\beta_{Zr}$ phase, along with $\omega_{iso}$ phase disappearance (Figure 7b).

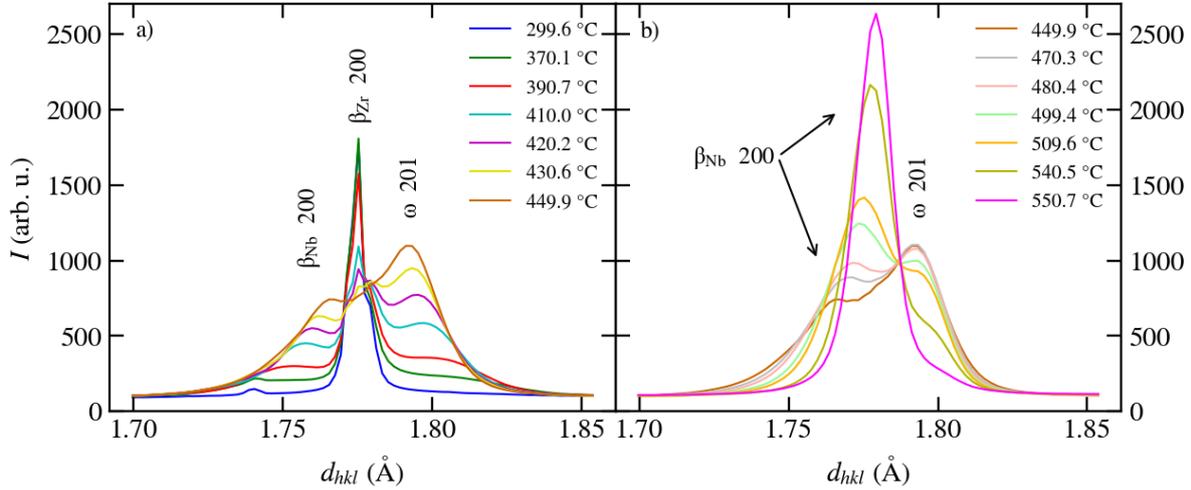

**Fig. 7:** XRD patterns showing the temperature dependence of the peaks in the vicinity of 2 0 0 diffraction peak of the β phase - a) temperature interval 300 – 450 °C, b) temperature interval 450 – 550 °C. The indexation of the peaks is also depicted

The 2 0 1 peak of the ω phase increases up to 450 °C (Figure 7a) while further increase of the temperature causes its decrease. Therefore, we may assume that in the whole temperature range 300 °C - 450 °C, the volume fraction of the ω phase increases and the decrease in electrical resistance cannot yet be explained by the dissolution of the ω phase. Our hypothesis is that during ω particle growth, the nature of the ω/β interfaces changes from coherent to semicoherent and consequently, elastic strains are partially released. This would lead to lower scattering of conduction electrons, and in turn to a lower electrical resistance.

### 4.4. Temperature range from 450 °C to 555 °C

The next stage of linear heating, approximately between 450 °C and 555 °C, is marked by an endothermic process, see the DSC curve in Figure 5. TEM micrographs of samples heated to 500 °C and 540 °C, Figures 6 c), d), show lower density of ω particles while the remaining ω particles increase in size. It can be observed that ω particles are cuboidal in the presently studied Zr15Nb alloy. This shape corresponds to a large lattice misfit between β and ω phases [46]. The misfit defined by Equation (2) is around 2 %, following from the lattice parameter values in Figure 3. In contrast to the ellipsoidal ω particles in Ti15Mo [47], similar cuboidal shape of ω particles was observed also in the Zr12Nb alloy [24].

Note that a significant increase in the size of ω particles correlated with the disappearance of the original $\beta_{Zr}$ phase around 430 °C (see Figure 3 or 4). Concurrently, with the disappearance of the $\beta_{Zr}$ phase, a small bump can be observed on the downward trend of the electrical resistance. A similar effect, although stronger, was observed in the Zr12Nb alloy [24]. Interestingly, a bump in the downward slope of the electrical resistance was also detected at analogous temperatures in the Ti15Mo alloy; however, the effect was much weaker [37]. The slower decrease of the electrical resistance around 450 °C means that the effect of ongoing microstructure changes which decrease electrical resistance is weaker around this temperature but still prevails over the phonon scattering contribution. Most likely, the ω phase dissolution, i.e. the gradual disappearance of ω/β interfaces which serve as obstacles for electron drift, is relatively slow at first. Between approximately 490 °C and 555 °C, the decrease of ω particles density is much faster, resulting in a faster decrease of electrical resistance.

The thermal expansion of the sample returns to an approximately linear trend in the discussed temperature range (450 °C - 555 °C). The steeper slope of thermal expansion curve is an effect of decreasing Nb enrichment of the $β_{Nb}$ phase (Nb diffuses back to regions of dissolved ω), which is visible in the relatively fast increase of $β_{Nb}$ lattice parameter and molar volume (Figures 3 and 4, respectively). The increasing lattice parameters of the $β_{Nb}$ phase is clearly depicted in Figure 7b as the related peak is shifted to higher interplanar distances.

Finally, at the temperature of 555 °C, all remaining ω phase dissolves and a rapid precipitation of the α phase occurs in the material as follows from the refinement of HEXRD data in Figures 3 and 4. All employed experimental techniques exhibit a strong response to this phase change. A sharp exothermic peak with the onset at 555 °C can be observed in the DSC signal (red curve in Figure 5). Sharp peaks are also present in the derivatives of electrical resistance and thermal expansion (gray and green lines in Figure 5, respectively). More detail investigation of DSC, electrical resistance, thermal expansion data and HEXRD data places the onset of this abrupt transition to 545 °C. Nevertheless, the α peaks below approximately 555 °C are very shallow and could not be picked up by the Le Bail refinement.

From Figure 8 it follows that the ω and α phases coexist in a short temperature range between approximately 545 °C and 555 °C. Once the α embryos start to form, ω phase dissolution is enhanced, resulting in an endothermic event in the DSC curve. The faster ω phase dissolution is also reflected in the steeper decline of the electric resistance curve due to disappearing ω/β interfaces. The steep increase in thermal expansion already in the range of coexistence of ω and α can be explained by a combination of two mechanisms. First, the α phase has a larger molar volume than both the ω and β phases (see Figure 4). Second, the ω phase, which is lean in Nb, dissolves in the $β_{Nb}$ matrix, lowering its average Nb content. Decreasing Nb content results in an increase of $β_{Nb}$ lattice parameter [19, 42] (see also the $β_{Nb}$ lattice parameter evolution in Figure 3), which contributes macroscopically to the fast thermal expansion (see Figure 5).

The coexistence of the β + ω + α phases just below the ω solvus was also observed in the Zr12Nb and Ti15Mo alloys [21, 28]. In Ti15Mo, it was shown using neutron diffraction that the coexistence of ω and α depends on the heating rate: ω and α were detected simultaneously during a slow heating (1.9 °C/min), while no coexistence was observed during a faster heating (5 °C/min) [48].

It is interesting to note that despite enhancing the ω dissolution by the presence of α embryos, the growth of α precipitates is still hindered by the remaining ω particles. Once the ω solvus is reached at approximately 555 °C and all remaining ω particles dissolve (see the selected HEXRD patterns in Figure 7), α phase starts to grow very rapidly, as evidenced by the sharp exothermic peak in the DSC and by a steep increase in electrical resistance (in which the effect of α precipitation adds to the general trend of phonon scattering).

### 4.5. Above the 555 °C

After the dissolution of the ω phase, significant changes in the lattice parameters of both α and β phases are observed, see Figure 3. The $a_α$ parameter decreases and the $c_α$ increases. Overall, the molar volume of the α phase decreases, see Figure 4. Simultaneously, both the β lattice parameter and the β molar volume (Figures 3 and 4, respectively) decrease as the Nb is rejected from the newly forming α particles into the β matrix. This behavior is evidenced in the HEXRD measurement (Figure 2) by a shift of the β peaks toward lower $d_{hkl}$ values. Between the ω solvus (555 °C) and the eutectoid temperature (620 °C), the equilibrium volume fraction of the α phase based on theoretical phase diagram (Figure 1) is significantly larger than that of the β phase resulting in rapid α phase growth. The TEM micrograph of a sample heated to 600 °C and quenched shows fine α precipitates in the β matrix (bright field TEM image in Figure 6e). The dark field image in Figure 6f) shows new small ω particles which formed in between α particles only during quenching.

Above the eutectoid temperature, the α phase starts to dissolve, since its equilibrium volume fraction abruptly becomes much lower, as follows from the phase diagram. The dissolution of the α phase is also accompanied by a "plateau" in the evolution of thermal expansion, see Figure 4. This means that the dissolution of the α phase with a higher molar volume prevails over the general thermal dilatation.

Although theoretical calculations (Thermo-Calc) predict the β transus temperature of Zr15Nb to be approximately 650 °C (see Figure 1), neither of the experiments was able to reliably detect it. Even at the highest temperature reached during the HEXRD measurement, 800 °C, peaks of the α phase are still visible. This discrepancy may be caused by the Thermo-Calc database overestimating the β stabilizing effect of Nb due to a relatively sparse dataset it is built upon (especially at higher Nb content). An inconsistency between the calculated phase diagram and experimental data was however reported even in Zr2.5Nb alloy [49]. Furthermore, it must be noted that the theoretical phase diagram does not account for the oxygen content in the alloy. Oxygen is known to be a strong α stabilizer in Zr-Nb alloys, an therefore its concentration of about 1000 wppm (the studied alloy, see the Experimental methods section) can raise the β transus temperature by a few tens of degrees [50–52]. Finally, the β transus cannot be unambiguously determined during the in-situ heating experiment, as the material is not in a true thermodynamic equilibrium and there might be simply not enough time for α phase dissolution.

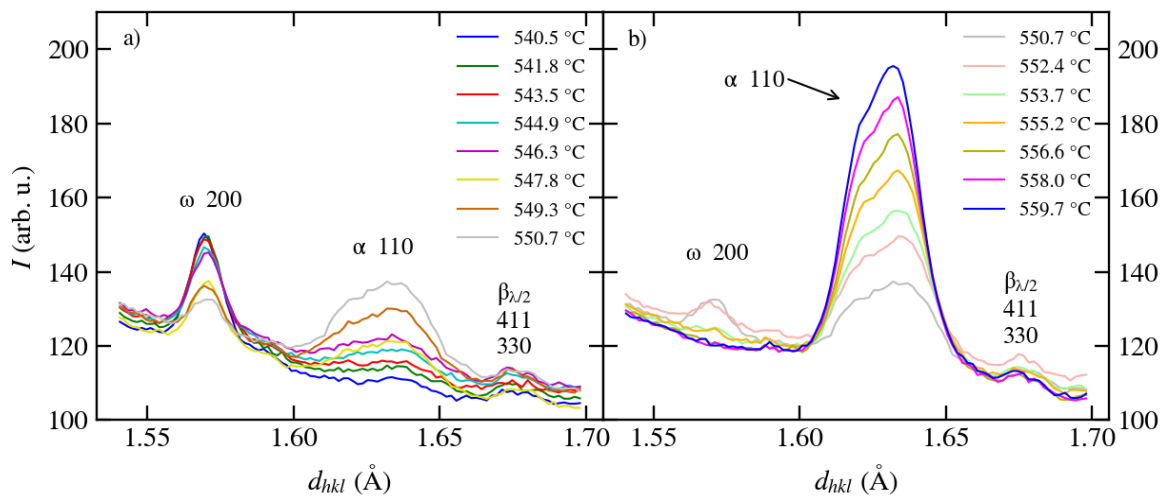

**Fig. 8:** Sequence of HEXRD patterns acquired at temperatures around the transformation near 555 °C. Selected peaks are indexed

   4.6. On the order of observed phase transformations

All phase transformations observed in this study – including the evolution and growth of the ω phase, subsequent ω dissolution, decomposition of the metastable β phase into $β_{Nb}$ and $β_{Zr}$, and nucleation and growth of the α phase – are classified as the first-order ones. The β → α transformation is a typical nucleation and growth process [53], consistent with the sharp exothermic peak observed in the heat-flow curve (Figure 5). The β → ω transformation is more complex, involving both a displacive collapse of adjacent $(111)_β$ planes and a diffusional rejection of solute atoms from the growing ω particles. The lattice-plane collapse itself has been demonstrated to be the first-order transformation [54]. However, the β → ω transformation is known to exhibit strong pre-transition phenomena, such as diffuse scattering and phonon softening, before discrete ω precipitates form [55]. Li et al. [56] suggested that the early stages of ω phase evolution are controlled by a second-order coherent spinodal decomposition of the β matrix. These observations suggest that the ω formation may be initiated by continuous, spinodal-like processes without a latent heat signature, but the transformation ultimately proceeds through a first-order mechanism. Our DSC results show exothermic and endothermic events associated with ω growth and dissolution, respectively, confirming the first-order nature of the transformation. Finally, the thermal signature of decomposition of the β phase into $β_{Zr}$ and $β_{Nb}$ cannot be unambiguously resolved in the DSC curve since it overlaps with ω growth. Nevertheless, the evolution of the XRD profiles (Figure 7) provides the clear evidence of the first-order behavior, with the appearance of a $β_{Nb}$ peak and coexistence of $β_{Zr}$ and $β_{Nb}$ over a finite temperature interval. This is consistent with β phase separation in Nb-Zr alloys [57].

4.7. Comparison of the Zr15Nb alloy with previously published Zr12Nb alloy

When compared to previously investigated Zr12Nb alloy [24], the Zr15Nb exhibits a similar phase transformation sequence, namely that ω phase and two β phases ($β_{Zr}$ and $β_{Nb}$ which are lean and rich in Nb, respectively) form in the material. However, several important differences can be found:

- The volume fraction of the $ω_{ath}$ phase in the initial condition (solution treated) is lower in Zr15Nb than in Zr12Nb. In fact, peaks corresponding to the $ω_{ath}$ were not detected by HEXRD at all, cf. Figure 2. Nevertheless, a certain amount of $ω_{ath}$ is present, as evidenced by TEM (Figure 6a) and by the initial decrease in electrical resistance associated with $ω_{ath}$ dissolution (Figure 5). Thus, an increased Nb content reduces the volume fraction of the $ω_{ath}$ phase in the solution treated condition.
- The size of the $ω_{ath}$ phase particles is significantly smaller in the Zr15Nb alloy in comparison to the Zr12Nb alloy, cf. Figure 6a) in this study and Figure 8a) in [24].
- Much stronger exothermic peak in the DSC curve in the temperature range of approximately 300 °C - 400 °C is observed in Zr15Nb alloy (cf. the red curves in Figure 5 in this study and in Figure 9 in [24]). Note, that we cannot claim whether the maximum volume fraction of the ω phase is larger in Zr15Nb compared to Zr12Nb, since the statistics of the HEXRD data were not sufficient to determine the volume fractions of individual phases due to the large grain size of the sample. From our data (strong exothermic DSC peak) we may only assume that the maximum amount of the ω particles relative to the amount in the initial condition is larger in the currently studied Zr15Nb.
- In Zr12Nb alloy, the maximum volume fraction of the ω phase was reached at the end of the second decrease in electrical resistance (around 430 °C in Zr12Nb). In the currently studied alloy, Zr15Nb, this point on the electrical resistance curve corresponds to approximately 450 °C. This statement was confirmed in detail by XRD experiments (see Figure 7).

- The temperature of $β_{Zr}$ phase dissolution is similar for both alloys (approximately 450 °C).

- The ω solvus temperature is slightly decreased by higher Nb content (560 °C and 555 °C in Zr12Nb and Zr15Nb, respectively).
- Finally, the evolution of the electrical resistance above the ω solvus suggests that in the previously studied Zr12Nb alloy, the amount of the α phase is significantly larger than in the present Zr15Nb. In the Zr12Nb alloy, a greater increase in electrical resistance related to α precipitation (approximately 560 °C - 630 °C) and a more distinct decrease connected to its dissolution (630 °C - 720 °C) were detected, cf. Figures 5 in this study and in [24]. Therefore, we may assume that the amount of α in Zr12Nb is larger than in the presently investigated Zr15Nb during the same heating procedure.

## 5. Conclusions

In this study, the phase transformation behavior of the Zr15Nb alloy was investigated through a combination of differential scanning calorimetry (DSC), electrical resistance, thermal expansion, high-energy X-ray diffraction (HEXRD) measurements and transmission electron microscopy (TEM).

The main findings of the investigation can be summarized as follows:

- The electrical resistance of the Zr15Nb alloy initially decreases from room temperature to approximately 130 °C, likely due to the release of elastic strains at β/ω interfaces, similar to observations in the Zr12Nb alloy.
- Between 220 °C and 420 °C, the alloy undergoes two exothermic processes that are related to chemical stabilization and growth of the ω phase, respectively, leading to a slower increase in thermal expansion (Nb enrichment of the β phase decreases its lattice parameter, which acts against thermal dilatation). The growth of ω particles eventually results in a decrease in electrical resistance as the ω/β interface becomes less coherent.
- A key finding is the observation of two bcc β phases: $β_{Zr}$ and $β_{Nb}$, whose lattice parameters have close values. As the temperature increases, Nb enrichment of the β matrix leads to the decomposition of $β_{Zr}$ and the formation of $β_{Nb}$. Additionally, changes in the β lattice parameter due to Nb redistribution influence both thermal expansion and electrical resistance.
- At temperatures between 420 °C and 555 °C, the ω phase dissolves and the α phase begins to precipitate. This transition is marked by sharp changes in DSC, electrical resistance, and thermal expansion, with evidence of α precipitation starting around 545 °C. Notably, the dissolution of the ω phase is enhanced by the presence of α embryos, while the rapid growth of the α phase starts only after all ω is dissolved. The α phase growth results in a significant increase in thermal expansion.
- The dissolution of the α phase begins around 600 °C, which corresponds to the eutectoid temperature.
- Our study also points to discrepancies between experimental data and theoretical predictions, particularly regarding the β transus temperature, which is higher than predicted by Thermo-Calc simulations. This deviation may be partly due to underestimation of the β-transus by Thermo-Calc or to the oxygen content in the alloy, which stabilizes the α phase.

In summary, this study provides a detailed understanding of the phase evolution, microstructural changes, and the corresponding changes in thermal and electrical properties of the Zr15Nb alloy. The interplay between the β, ω, and α phases at different temperatures offers valuable insights into the behavior of the alloy and the mechanisms governing phase transformations, with implications for future materials design and processing strategies. The study also shows the advantage of using laboratory in-situ measurement methods concurrently with in-situ HEXRD and ex-situ TEM.

**Acknowledgements:**

Financial support by the Czech Science Foundation under the Project No. 24-10512S is gratefully acknowledged. Financial support by the Operational Programme Johannes Amos Comenius of the MEYS of the Czech Republic, within the frame of the project Ferroic Multifunctionalities (FerrMion) [project No. CZ.02.01.01/00/22_008/0004591], co-funded by the European Union is also gratefully acknowledged. A.V. acknowledges Grant Agency of the Charles University, project no. 412522. The Deutsches Elektronen-Synchrotron (DESY) is acknowledged for the provision of synchrotron radiation facilities in the framework of the I-20230198 EC proposal. P.B.-V. acknowledges financial support from the Spanish Ministry of Science (MCIN) through the Ramón y Cajal grant RYC2020-029585-I and the project PID2022-141670OA-I00. The Agency for Management of University and Research Grants (AGAUR) is acknowledged for the financial support in the project 2023PROD00078. The European Commission is acknowledged for funding support in the project LIGHTFORGE under grant agreement 101112392. J. P. acknowledges the MGML facilities (mgml.eu, supported within the program of Czech Research Infrastructures under the Project No. LM2023065).

**Data availability:**

The data that support the findings of this study are openly available in the Zenodo repository at https://doi.org/10.5281/zenodo.17590045 under the CC-BY 4.0 license.

**Supplementary information:**

'Not applicable'

**Ethical approval:**

'Not applicable', because this research does not contain any studies with humans or animals.

**References:**

1. B. Cox, in *Uhlig's Corrosion Handbook* (John Wiley & Sons, Ltd, 2011), pp. 893–900.

2. V.P. Mantripragada, B. Lecka-Czernik, N.A. Ebraheim, A.C. Jayasuriya, *J. Biomed. Mater. Res. A* **101**, 3349 (2013). https://doi.org/10.1002/jbm.a.34605

3. H.G. Rickover, L.D. Geiger, B. Lustman, *History of the Development of Zirconium Alloys for Use in Nuclear Reactors* (Energy Research and Development Administration, Washington, D.C., 1975).

4. A.M. Weinberg, *Phys. Today* **46**, 70 (1993). https://doi.org/10.1063/1.2809022

5. F.Y. Zhou, B.L. Wang, K.J. Qiu, et al., *J. Biomed. Mater. Res. B Appl. Biomater.* **101B**, 237 (2013). https://doi.org/10.1002/jbm.b.32833

6. Z. Duan, H. Yang, Y. Satoh, et al., *Nucl. Eng. Des.* **316**, 131 (2017). https://doi.org/10.1016/j.nucengdes.2017.02.031

7. C. Lemaignan, A.T. Motta, in *Materials Science and Technology* (John Wiley & Sons, Ltd, 2006)

8. C.L. Whitmarsh, *REVIEW OF ZIRCALOY-2 AND ZIRCALOY-4 PROPERTIES RELEVANT TO N.S. SAVANNAH REACTOR DESIGN* (1962)

9. Mrs. I.T. Bethune, C.D. Williams, *J. Nucl. Mater.* **29**, 129 (1969). https://doi.org/10.1016/0022-3115(69)90134-2

10. G. Sabol, *J. ASTM Int.* **2** (2005). https://doi.org/10.1520/JAI12942

11. A.V. Nikulina, *Met. Sci. Heat Treat.* **46**, 458 (2004). https://doi.org/10.1007/s11041-005-0002-x

12. A. Garde, R. Comstock, G. Pan, et al., *Zircon. Nucl. Ind. 16th Int. Symp.* (2012). https://doi.org/10.1520/STP152920120031zheng

13. R. Kondo, N. Nomura, H. Doi, et al., *Mater. Trans.* advpub (2016). https://doi.org/10.2320/matertrans.MI201512

14. B.A. Cheadle, S.A. Aldridge, *J. Nucl. Mater.* **47**, 255 (1973). https://doi.org/10.1016/0022-3115(73)90109-8

15. L. Nie, Y. Zhan, T. Hu, et al., *J. Mech. Behav. Biomed. Mater.* **29**, 1 (2014). https://doi.org/10.1016/j.jmbbm.2013.08.019

16. L. Nie, Y. Zhan, H. Liu, C. Tang, *Mater. Des.* **53**, 8 (2014). https://doi.org/10.1016/j.matdes.2013.07.008

17. P. Chui, *Vacuum* **143**, 54 (2017). https://doi.org/10.1016/j.vacuum.2017.05.039

18. C.R.F. Azevedo, *Eng. Fail. Anal.* **18**, 1943 (2011). https://doi.org/10.1016/j.engfailanal.2011.06.010

19. R. Kondo, N. Nomura, Suyalatu, et al., *Acta Biomater.* **7**, 4278 (2011). https://doi.org/10.1016/j.actbio.2011.07.020

20. N. Nomura, Y. Tanaka, Suyalatu, et al., *Mater. Trans.* **50**, 2466 (2009). https://doi.org/10.2320/matertrans.M2009187

21. M. Bönisch, M. Calin, T. Waitz, et al., *Sci. Technol. Adv. Mater.* **14**, 055004 (2013). https://doi.org/10.1088/1468-6996/14/5/055004

22. H.Y. Kim, Y. Ikehara, J.I. Kim, et al., *Acta Mater.* **54**, 2419 (2006). https://doi.org/10.1016/j.actamat.2006.01.019

23. Y. Mantani, M. Tajima, *Mater. Sci. Eng. A* **442**, 409 (2006). https://doi.org/10.1016/j.msea.2006.03.124

24. A. Veverková, P. Harcuba, J. Veselý, et al., *J. Mater. Res. Technol.* **23**, 5260 (2023). https://doi.org/10.1016/j.jmrt.2023.02.076

25. J. Šmilauerová, P. Harcuba, J. Stráský, et al., *Acta Mater.* **81**, 71 (2014). https://doi.org/10.1016/j.actamat.2014.06.042

26. D. De Fontaine, N.E. Paton, J.C. Williams, *Acta Metall.* **19**, 1153 (1971). https://doi.org/10.1016/0001-6160(71)90047-2

27. A. Devaraj, S. Nag, R. Srinivasan, et al., *Acta Mater.* **60**, 596 (2012). https://doi.org/10.1016/j.actamat.2011.10.008

28. K. Bartha, J. Stráský, P. Barriobero-Vila, et al., *J. Alloys Compd.* **867**, 159027 (2021). https://doi.org/10.1016/j.jallcom.2021.159027

29. P. Barriobero-Vila, G. Requena, F. Warchomicka, et al., *J. Mater. Sci.* **50** (2014).
https://doi.org/10.1007/s10853-014-8701-6

30. P. Barriobero-Vila, G. Requena, S. Schwarz, et al., *Acta Mater.* **95**, 90 (2015).
https://doi.org/10.1016/j.actamat.2015.05.008

31. E. Aeby-Gautier, A. Settefrati, F. Bruneseaux, et al., *J. Alloys Compd.* **577**, S439 (2013).
https://doi.org/10.1016/j.jallcom.2012.02.046

32. N. Schell, A. King, F. Beckmann, et al., *Mater. Sci. Forum* **772**, 57 (2014).
https://doi.org/10.4028/www.scientific.net/MSF.772.57

33. P. Staron, F. Beckmann, T. Lippmann, et al., *Mater. Sci. Forum* **690**, 192 (2011).
https://doi.org/10.4028/www.scientific.net/MSF.690.192

34. A.P. Hammersley, *J. Appl. Crystallogr.* **49**, 646 (2016).
https://doi.org/10.1107/S1600576716000455

35. J. Rodríguez-Carvajal, *Phys. B Condens. Matter* **192**, 55 (1993). https://doi.org/10.1016/0921-4526(93)90108-I

36. J. Šmilauerová, P. Harcuba, D. Kriegner, V. Holý, *J. Appl. Crystallogr.* **50**, 283 (2017).
https://doi.org/10.1107/S1600576716020458

37. P. Zháňal, P. Harcuba, J. Šmilauerová, et al., *Acta Phys. Pol. A* **128**, 779 (2015).
https://doi.org/10.12693/APhysPolA.128.779

38. P. Zháňal, K. Václavová, B. Hadzima, et al., *Mater. Sci. Eng. A* C, 886 (2016).
https://doi.org/10.1016/j.msea.2015.11.029

39. P.L. Rossiter, *The Electrical Resistivity of Metals and Alloys* (Cambridge University Press, Cambridge, 1987)

40. P. Zháňal, P. Harcuba, M. Hájek, et al., *J. Appl. Crystallogr.* **52**, 1061 (2019).
https://doi.org/10.1107/S1600576719010537

41. P. Zháňal, P. Harcuba, M. Hájek, et al., *Mater. Sci. Eng. A* **760**, 28 (2019).
https://doi.org/10.1016/j.msea.2019.05.034

42. D. De Fontaine, *J. Less Common Met.* **19**, 271 (1969). https://doi.org/10.1016/0022-5088(69)90102-5

43. A. Kumar, B. Velayutham, S. Chandrasekaran, et al., *Metall. Mater. Trans. A* **50**, 1423 (2019).
https://doi.org/10.1007/s11661-019-05184-4

44. N. Schell, F. Beckmann, S. Wiegand, et al., *Mater. Sci. Eng. A* **615**, 117 (2014).
https://doi.org/10.1016/j.msea.2014.07.061

45. P. Staron, T. Lippmann, F. Beckmann, et al., *Mater. Sci. Eng. A* **527**, 3670 (2010).
https://doi.org/10.1016/j.msea.2010.03.062

46. J. Šmilauerová, P. Harcuba, J. Stráský, et al., *Acta Mater.* **60**, 439 (2012).
https://doi.org/10.1016/j.actamat.2011.10.033

47. A. Garde, R. Comstock, G. Pan, et al., *J. Nucl. Mater.* **431**, 142 (2012).
https://doi.org/10.1016/j.jnucmat.2012.05.003

48. R. Kondo, N. Nomura, Suyalatu, et al., *Mater. Trans.* **55**, 1012 (2014).
    https://doi.org/10.2320/matertrans.M2013164

49. M. Bönisch, M. Calin, T. Waitz, et al., *Scr. Mater.* **68**, 553 (2013).
    https://doi.org/10.1016/j.scriptamat.2012.11.008

50. H.Y. Kim, Y. Ikehara, J.I. Kim, et al., *Acta Mater.* **56**, 2421 (2008).
    https://doi.org/10.1016/j.actamat.2008.01.008

51. Y. Mantani, M. Tajima, *Mater. Sci. Eng. A* **442**, 417 (2006).
    https://doi.org/10.1016/j.msea.2006.03.123

52. A. Veverková, P. Harcuba, J. Veselý, et al., *J. Mater. Res. Technol.* **23**, 5275 (2023).
    https://doi.org/10.1016/j.jmrt.2023.03.021

53. J. Šmilauerová, P. Harcuba, J. Stráský, et al., *Acta Mater.* **82**, 88 (2014).
    https://doi.org/10.1016/j.actamat.2014.06.055

54. D. De Fontaine, N.E. Paton, J.C. Williams, *Acta Metall.* **20**, 1165 (1972).
    https://doi.org/10.1016/0001-6160(72)90077-1

55. A. Devaraj, S. Nag, R. Srinivasan, et al., *Acta Mater.* **61**, 502 (2013).
    https://doi.org/10.1016/j.actamat.2012.09.013

56. K. Bartha, J. Stráský, P. Barriobero-Vila, et al., *J. Alloys Compd.* **872**, 159130 (2021).
    https://doi.org/10.1016/j.jallcom.2021.159130

57. P. Barriobero-Vila, G. Requena, F. Warchomicka, et al., *J. Mater. Sci.* **51**, 3501 (2016).
    https://doi.org/10.1007/s10853-016-9827-1